\documentclass[showpacs,preprintnumbers,amsmath,amssymb]{revtex4}

\usepackage{graphicx}
\usepackage{dcolumn}
\usepackage{bm}
\usepackage{color}
\usepackage{ulem}
\usepackage[T1]{fontenc}

\newcommand{\aap}{\rm Astron.~\&~Astrophys.~}

\newcommand{\apjs}{\rm Astrophys.~J.~Supp.~}

\newcommand{\physrep}{\rm Phys.~Rep.~}

\begin{document}

\preprint{APS/123-QED}

\title{Cosmological solutions to the Lithium problem: Big-bang nucleosynthesis with photon cooling, $X$-particle decay and a primordial magnetic field}

\author{Dai G. Yamazaki$^{1}$}
 \email{yamazaki.dai@nao.ac.jp}
\author{Motohiko Kusakabe$^{2,3}$}%
\author{Toshitaka Kajino$^{1,4}$}
\author{Grant. J. Mathews$^{1,5}$}
\author{Myung-Ki Cheoun$^{3}$}
\affiliation{%
$^{1}$National Astronomical Observatory of Japan, Mitaka, Tokyo 181-8588, Japan}%
\affiliation{%
$^{2}$School of Liberal Arts and Science, Korea Aerospace University, Goyang 412-791, Korea}%
\affiliation{%
$^{3}$Department of Physics, Soongsil University, Seoul 156-743, Korea}%
\affiliation{%
$^{4}$Department of Astronomy, Graduate School of Science, The University of Tokyo,
Hongo 7-3-1, Bunkyo-ku, Tokyo 113-0033, Japan}%
\affiliation{%
$^{5}$Center for Astrophysics, Department of Physics, University of Notre Dame, Notre Dame, Indiana 46556, USA}%
\date{\today}

\begin{abstract}
The $^7$Li abundance calculated in BBN with the baryon-to-photon ratio fixed from fits to the CMB power spectrum is inconsistent with the observed lithium abundances on the surface of metal-poor halo stars. Previous cosmological solutions proposed to resolve this $^7$Li problem include photon cooling (possibly via the Bose-Einstein condensation of a scalar particle) or the decay of a long-lived $X-$particle (possibly the next-to-lightest supersymmetric particle). In this paper we reanalyze these solutions, both separately and in concert. We also introduce the possibility of a primordial magnetic field (PMF) into these models. We constrain the $X-$particles and the PMF parameters by the observed light element abundances using a likelihood analysis to show that the inclusion of all three possibilities leads to an optimum solution to the lithium problem. We deduce allowed ranges for the $X-$particle parameters and energy density in the PMF that can solve $^7$Li problem.
\end{abstract}

\maketitle
\section{\label{sec:introduction}Introduction}
The primordial $^7$Li abundance is derived from observations of low-metallicity halo stars \cite{2010AandA...522A..26S}. At the same time one can theoretically estimate the
$^7$Li abundance from big bang nucleosynthesis (BBN) by fixing the baryon-to-photon ratio ($\eta$) to the value determined from fits to the power spectrum of the cosmic microwave background (CMB) \cite{WMAP_9yr_Arxiv}. Nevertheless, there still remains a gap between astronomically observed and cosmologically inferred $^7$Li abundances. This conundrum is called the ``Li problem'' \cite{2011ARNPS..61...47F}. In this work we consider three possible resolutions of this conundrum both separately and in concert as a means to deduce the optimum parameter space to resolve this dilemma.

Table \ref{tab:models} summarizes the status and weakness of each of the models and combinations considered here. The first and second rows show models and corresponding $\eta$ values, respectively. Below the third row, the first and second columns show nuclides and observational constraints on primordial abundances, respectively, while the third to seventh columns correspond to calculated results in models of (i) standard BBN (SBBN), (ii) photon-cooling after the BBN epoch, (iii) BBN with a long-lived radiatively decaying $X-$particle with the parameter set ($\tau_X$, $\zeta_X$)=($10^5$ s, $2\times 10^{-10}$ GeV), (iv) BBN with a magnetic field, (v) BBN with photon cooling, $X-$particles with the parameter set ($\tau_X$, $\zeta_X$)=($10^6$ s, $2\times 10^{-10}$ GeV) \cite{2013PhLB..718..704K}, and a magnetic field, respectively. Check marks ($\checkmark$) or hyphens (-) indicate that calculated abundances are consistent or inconsistent, respectively, with observations. A question mark (?) indicates that a calculated result is near (though slightly outside of) the observationally allowed region. In the last row, ``high'' indicates that the resulting $^6$Li abundance can be higher than in the SBBN but within the current observational upper limit \cite{2013AA...554A..96L}.

As a first possibility we consider photon cooling. It is one of the more innovative solutions to the Li problem. 
If photons are cooled sometime after the end of BBN, then the baryon-to-photon ratio during BBN is smaller than the present value and the excess production of $^7$Li can be avoided.
The candidate mechanisms for such cooling are a possible scalar particle that forms a Bose-Einstein condensate (BEC) during the post BBN epoch, but before photon decoupling. 
Although the dark-matter axion was initially considered as a candidate particle for this BEC \cite{2009PhRvL.103k1301S,2012PhRvD..85f3520E}, 
the cooling occurs too late so that it disagrees with the CMB constraints on the chemical potential $\mu$ and the Compton $y$ parameter (see Sec. \ref{sec:distortion}).
Another possibility, however could be resonant oscillations between photons and light Abelian gauge bosons in the hidden sector \cite{Jaeckel:2008fi}.

Another proposed solution to the Li problem is the radiative decay of a hypothetical exotic long-lived ($X$) particle after BBN \cite{2013PhLB..718..704K}. 
As noted in Table \ref{tab:models}, however, the decaying particle model (without assuming photon cooling) cannot resolve the $^7$Li problem by itself~\cite{Ellis:2005ii,Kusakabe:2006hc}. Here we point out, however, that this paradigm could solve the problem of deuterium overproduction in the photon cooling model \cite{2013PhLB..718..704K}. In such a hybrid model consisting of photon cooling plus a radiatively decaying particle, the primordial D abundance is decreased via the photodisintegration reaction, $^2$H($\gamma$, $n$)$^1$H, by nonthermal photons produced from the decaying $X-$particle \cite{Lindley1979MNRAS.188P..15L,Ellis:1984er,Dimopoulos:1987fz,Reno:1987qw,Terasawa:1988my,Kawasaki:1994af,Kawasaki:1994sc,Jedamzik:1999di,Kawasaki:2000qr,Cyburt:2002uv,Kawasaki:2004qu,Jedamzik:2004er,Jedamzik:2004ip,Ellis:2005ii,Jedamzik:2006xz,Kusakabe:2006hc,Kusakabe:2008kf,Cyburt:2010vz,Pospelov:2010cw}.

Of relevance to the present work is the fact that the cosmic expansion rate can also be boosted by the energy density of a primordial magnetic field (PMF). Hence, BBN with a PMF leads to different results than without a PMF \cite{2012arXiv1204.6164K,2012PhRvD..86l3006Y} and can potentially improve the solution to the Li problem.
In this paper, therefore, we consider the combined effects of photon cooling, the radiative decay of an $X-$particle, and a PMF on BBN. 
We then utilize a maximum likelihood analysis of the observed abundances of light elements up to Li to constrain the parameters characterizing the $X-$particles and the PMF.

We also note that another possible solution to the Li problem has been proposed based upon a variant of the decaying $X-$particle scenario. This
solution is based upon exotic atomic and nuclear reactions induced by a long-lived negatively charged massive particle $X^-$ \cite{Cahn:1980ss} during BBN \cite{Dimopoulos:1989hk,DeRujula:1989fe,Pospelov:2006sc,Kohri:2006cn,Cyburt:2006uv,Hamaguchi:2007mp,Bird:2007ge,Kusakabe:2007fu,Kusakabe:2007fv,Jedamzik:2007cp,Jedamzik:2007qk,Kamimura:2008fx,Pospelov:2007js,Kawasaki:2007xb,Jittoh:2007fr,Jittoh:2008eq,Jittoh:2010wh,Pospelov:2008ta,Kamimura2010,Kusakabe:2010cb,Pospelov:2010hj,Cyburt:2012kp,2012PhRvC..85d4602D,Kusakabe:2013tra,2013PhRvD..88h9904K,Konishi:2013gda}. Such a charged particle with mass $m_X \gg {\cal O}$(1 GeV) can recombine with positively charged nuclei via radiative-capture reactions of bare nuclei and $X^-$ \cite{Dimopoulos:1989hk,DeRujula:1989fe} or nuclear charge-exchange reactions between electronic ions and the $X^-$, especially $^7$Be$^{3+} + X^- \rightarrow ^7$Be$_X + e^-$ \cite{Kusakabe:2013tra,2013PhRvD..88h9904K} in a late epoch of BBN. Bound states of nuclei and $X^-$, i.e., $A_X$ or $X$ nuclei, then induce atomic and nuclear reactions. Among the reactions, new types of resonant nuclear reactions of $^7$Be$_X+p \rightarrow ^8$B$_X +\gamma$ are made possible through an atomic excited state, $^8$B$_X^{\ast{\rm a}}$ \cite{Bird:2007ge} and an atomic ground state consisting of the $1^+$ nuclear excited state of $^8$B and an $X^-$, i.e., $^8$B$^\ast$($1^+$,0.770~MeV)$_X$~\cite{Kusakabe:2007fu}. Primordial $^7$Be nuclei are destroyed via these reactions, and the final abundance ratio of $^7$Li/H is reduced. Although this remains as an interesting solution to the Li problem, the destruction of deuterium does not occur in this model.
However, deuterium destruction is an essential part of models with photon cooling. Hence, we do not consider the $X^-$ paradigm in the models discussed below.

This paper is then organized as follows: In Sec. \ref{sec:models} we present details of each model and in Sec. \ref{sec:modelsC} we introduce a model that includes the combined effects on BBN from photon cooling, radiative $X$ decay, and a PMF.
In Sec. \ref{sec:distortion} we describe the constraints on the photon cooling from an analysis of the energy spectrum of the CMB. In Sec. \ref{sec:results} we show results of the BBN calculation and parameter search. We then discuss the effective number of relativistic degrees of freedom, and summarize this work.
\section{\label{sec:models}Model}
We have modified the Kawano BBN code \cite{Kawano_Code1992,1993ApJS...85..219S} to take into account the effects of (i) photon cooling (ii) the radiative decay of the $X-$particle, and (iii) a PMF. We assume that the photon cooling occurs after both BBN and the $X$ decay. The effect of the $X$ decay is included using a method similar to that described in Ref.~\cite{2013PhLB..718..704K}. The effect of photon cooling on the nuclear abundances enters through the difference between the baryon-to-photon ratio at the epoch of BBN and the value at the time of cosmological recombination. All relevant effects of a PMF are included as described in Ref.~\cite{2012arXiv1204.6164K}.
The reaction rates are taken from the JINA
REACLIB Database V1.0 \cite{2010ApJS..189..240C}.  We adopt a neutron life time of 
$\tau_\mathrm{n} = 878.5 (\pm 0.7_\mathrm{st} \pm 0.3 _\mathrm{sy})$ s 
\cite{2010PhRvC..82c5501S}. 
For nonthermal nucleosynthesis we have utilized updated reaction rates for $^4$He photodisintegration~\cite{Kusakabe:2008kf} derived from the cross section data obtained from precise measurements with laser-Compton photons~\cite{Shima:2005ix,Kii:2005rk}.

In Ref. \cite{2013PhLB..718..704K} we utilized the nonthermal nuclear transfer functions \cite{Kusakabe:2008kf} for a value of $\eta = 6.1\times 10^{-10}$ from the WMAP 3yr \cite{Spergel:2003cb} analyses obtained. However, the transfer functions depend upon the baryon-to-photon ratio, and should be corrected for changes in both the nuclear abundances and the $\eta$ value. In this study, therefore, transfer functions were updated by a self-consistent calculation  for BBN as a function of $\eta$. The energy loss rate for nonthermal photons and the rates for energy loss and the destruction of nonthermal nuclei produced by the energetic photons were thus modified. As long as the effects of the radiative decay of the $X$ are not too large and the light element abundances are not significantly different from the SBBN predictions, then consistent results could be derived in this model calculation. We calculated nuclear transfer functions in this nonthermal nucleosynthesis model for 11 different $\eta$ values. We then interpolated the transfer functions as a function of $\eta$ to deduce calculated abundances.

We then compare light element abundances calculated in the BBN model to observational abundance constraints.
We utilized element abundances constrained by observations as summarized on the second column of Table \ref{tab:models} [see Ref. \cite{2012PhRvD..86l3006Y} for details].
We adopt constraints on the baryon-to-photon number ratio as deduced from the WMAP 9yr \cite{WMAP_9yr_Arxiv} analysis,
\begin{eqnarray}
\eta = \frac{n_b}{n_\gamma}
=2.734 \times 10^{-8}~
\Omega_b h^2
=
(6.19 \pm 0.14) \times 10^{-10},
\label{baryon_to_photon_ratio}
\end{eqnarray}
where $n_b$ and $n_\gamma$ are the number densities of baryons and photons, respectively,
$\Omega_b$ is the baryon contribution to the closure density,
and $h$ is the Hubble parameter in units of 100 km s$^{-1}$ Mpc$^{-1}$.
We use $N_\nu = 3$ as the number of neutrino species.
This is not to be confused with the effective number of neutrino species $N_\mathrm{eff}$ discussed below.
We also note that the degeneracy between $\eta$ and $N_\mathrm{eff}$ is negligible based upon the results of the WMAP 9yr analysis \cite{WMAP_9yr_Arxiv} as shown in Table \ref{tabel2}.

\subsection{\label{sec:modelsA}Photon cooling}
When there is photon cooling the number density of photons at the BBN epoch is greater than that without it. Therefore, the baryon-to-photon ratio at the BBN epoch, $\eta_\mathrm{BBN}$ will be less than the value fixed by the CMB power spectrum. 
The predicted baryon-to-photon ratio in the photon cooling BEC model \cite{2012PhRvD..85f3520E}
is smaller due to the statistical degrees of freedom
by a factor of 
$(2/3)^{3/4}$ at BBN:
$\eta_\mathrm{BBN} = (4.57 \pm 0.10) \times 10^{-10}$ independently of the cooling epoch.
In the photon cooling model, the D, $^3$He and $^6$Li abundances increase, while the $^4$He and $^7$Li abundances decrease.
The $^7$Li abundance derived via a photon-cooling model can be consistent with the observed value. However, this leads to an overabundance of D. 

\subsection{\label{sec:modelsB}$X$ particle model}
Nonthermal electromagnetic energy injection caused by the radiative decay of a long-lived massive $X$ particle after the BBN epoch can generate nonthermal photons \cite{Lindley1979MNRAS.188P..15L,Ellis:1984er,Kawasaki:1994af,Kawasaki:1994sc}. Such decaying particles are predicted in particle theories beyond the standard model \cite{Feng:2003xh,Feng:2003uy}. Examples include the next-to-lightest supersymmetric particle. Energetic photons emitted by the radiative decay can react with background photons to create $e^+e^-$ pairs, and an electromagnetic cascade shower develops. The spectrum of nonthermal photons associated with this shower has an upper cutoff at energy $E_{\rm C}=m_e^2/22T$ with $m_e$ the electron mass~\cite{Kawasaki:1994sc}. At high temperature in the early universe, the cutoff energies are smaller.

As the temperature of the universe decreases over time, the nonthermal photons from the particle decay gradually disintegrate the light nuclei produced in BBN \cite{Lindley1979MNRAS.188P..15L,Ellis:1984er,Cyburt:2002uv,Kawasaki:2004qu,Jedamzik:2006xz,Kusakabe:2006hc}. Among such light nuclei, $^7$Be and D are the most strongly photodisintegrated because the threshold energies for $^7$Be and D photodisintegration are 1.5866 and 2.2246 MeV, respectively. They are mainly photodisintegrated via the $^7$Be+$\gamma\rightarrow ^3$He+$^4$He and D+$\gamma \rightarrow n +p$ reactions, where $\gamma$'s represent nonthermal photons.
Since these photodisintegrations begin when the cutoff energies exceed the threshold energies, the destruction of $^7$Be and D starts early ($ T \gtrsim 10^{7}$K) which corresponds to $\tau_X\lesssim 10^6$~s.
Therefore, if the cutoff energy is above the threshold energies for $^7$Be and D destruction, but below that of other nuclei, $^7$Be and D are destroyed by the nonthermal photon field, while other nuclei are unaffected.
Long after the BBN epoch, $^7$Be nuclei recombine with electrons, and are converted to $^7$Li via electron capture, i.e., $^7$Be + e$^-$ $\rightarrow$ $^7$Li + $\nu_\mathrm{e}$. The primordial $^7$Li abundance is, therefore, the sum of the abundances of $^7$Li and $^7$Be produced in BBN. The $^7$Be photodisintegration triggered by the $X$ decay thus reduces the primordial $^7$Li abundance.
One problem with the photon cooling model was D overproduction (see Table \ref{tab:models}). This can be fixed, however, by D photodisintegration in a BBN model in which both  photon cooling and a radiatively decaying particle are included. 

There are two parameters which characterize effects of the electromagnetic energy injection. One is $\zeta_X=(n_X^0/n_\gamma^0)E_{\gamma 0}$,
where $(n_X^0/n_\gamma^0)$ is the ratio of the number density of decaying $X$ particles and the number density of the background radiation before the decay of the $X$, 
and $E_{\gamma 0}$ is the energy of the photon emitted by the radiative decay.
The other parameter is $\tau_X$, the decay lifetime of the $X$ particle \cite{2013PhLB..718..704K}.
\subsection{\label{sec:modelsC}PMF model}
A third cosmological scenario considered here is that of a primordial magnetic field \cite{Yamazaki:2004vq,Yamazaki:2006bq,Yamazaki:2007oc,2010PhRvD..81b3008Y,2012PhR...517..141Y,2014PhRvD..89j3528Y}. 
When one adds the energy density of a possible PMF during the BBN epoch, the rate of the cosmic expansion is more rapid than without a PMF. 
In this case, the freeze-out of the weak reactions can occur earlier. The neutron abundance at the epoch of weak-reaction freeze-out then increases. 
 Due to the faster cosmic expansion the time interval after the freeze-out until $^4$He production also becomes shorter. 
Therefore, more neutrons survive the $\beta$-decay until the start of the $^4$He production epoch. 
As a result
\cite{2012arXiv1204.6164K,2012PhRvD..86l3006Y}
the $^4$He abundance increases significantly, while the D and $^3$He abundances increase moderately, 
 the $^6$Li abundance increases slightly, and the $^7$Li abundance decreases.
\subsection{\label{sec:modelsD}Likelihood analysis}
We have utilized a likelihood analysis to constrain the parameters of the $X$ particle and a PMF. The likelihood function for any observed value $A_\mathrm{obs}$ can be expressed as 
\begin{eqnarray}
L_A (\tau_X, \zeta_X, \rho_B) = 
\frac{1}{\sqrt{\mathstrut 2\pi}\sigma_{A_\mathrm{obs}}}
\exp 
\left[
  \frac{-(A_\mathrm{th}(\tau_X, \zeta_X, \rho_B)-A_\mathrm{obs})^2}{2\sigma^2_{A_\mathrm{obs}}}
\right],
\label{L_fun1st}
\end{eqnarray}
where $A_\mathrm{th}(\tau_X, \zeta_X, \rho_B)$ is the theoretical value calculated as a function of  $\eta$, and $A_\mathrm{obs}$ is the observed distribution from Table \ref{tab:models}.
To constrain $(\tau_X, \zeta_X, \rho_B)$, we use the combined likelihood functions as follows:
\begin{eqnarray}
L_\mathrm{LE}(\tau_X, \zeta_X, \rho_B) =
 L_\mathrm{D} ~ \times ~ 
 L_{^3\mathrm{He}} ~ \times ~
 L_{^4\mathrm{He}} ~ \times ~
 L_{^6\mathrm{Li}} ~ \times ~
 L_{^7\mathrm{Li}}~,
\label{L_fun2nd}
\end{eqnarray}
where the likelihood functions for respective parameters are defined by
\begin{eqnarray}
L_\mathrm{LE}(\tau_X) = 
\int\int d\zeta_X d\rho_B 
L_\mathrm{LE}(\tau_X, \zeta_X, \rho_B),
\label{L_funtau}
\\
L_\mathrm{LE}(\zeta_X) = 
\int\int d\tau_X d\rho_B 
L_\mathrm{LE}(\tau_X, \zeta_X, \rho_B),
\label{L_funzeta}
\\
L_\mathrm{LE}(\rho_B) = 
\int\int d\zeta_X d\tau_X 
L_\mathrm{LE}(\tau_X, \zeta_X, \rho_B).
\label{L_funpmf}
\end{eqnarray}
\section{\label{sec:distortion}Constraints from the CMB energy spectrum}
We assume that there is a cooling of the cosmic background radiation (CBR) between the BBN epoch and the epoch of cosmological recombination. Proposed cooling mechanisms include energy transfer to a light Abelian gauge boson that mixes with the photon \cite{Jaeckel:2008fi}, or the energy transfer to an axion \cite{2012PhRvD..85f3520E}. 
Although photon cooling via a dark-matter axion interaction is ruled out by CMB observations as shown below, 
our consideration is applicable to any general photon cooling mechanism such as a photon mixing with an additional gauge boson \cite{Jaeckel:2008fi}.

However, even in this case, the energy transfer to and from the CBR is constrained by the observed energy spectrum of the CMB. 
Since the energy transfer during a late epoch of the universe would distort the CMB blackbody spectrum, this possibility is
constrained by the observed consistency of the CMB spectrum with that of a perfect  blackbody
~\cite{Hu:1993gc,Feng:2003xh}. For epochs earlier than $z\sim10^7$,
thermal bremsstrahlung, [i.e. free-free emission ($eN\rightarrow eN\gamma$),
where $N$ is any ion] and radiative-Compton scattering ($e^-\gamma\rightarrow e^-\gamma\gamma$) act effectively to erase any
distortion of the CBR spectrum from that of a blackbody. For energy transfer in
later epochs $10^5<z<10^7$, processes changing the photon number become ineffective,
so that Compton scattering ($\gamma e^-\rightarrow \gamma e^-$) causes
the photons and electrons to achieve statistical (but not
thermodynamic) equilibrium. Then, the photon spectrum obeys a
Bose-Einstein distribution
\begin{eqnarray}
 f_\gamma(\vec{p}_\gamma)=\frac{1}{e^{\epsilon_\gamma/T+\mu}-1}~~,
 \label{distortion1}
\end{eqnarray}
where
$\vec{p}_\gamma$ and $\epsilon_\gamma$ are the momentum and energy of the photon, and 
$\mu$ is the dimensionless chemical potential derived from the
conservation of photon number. 

Analyses of the CMB data suggest a relatively low baryon density so that
double Compton scattering dominates the thermalization process. For 
a small value of $\mu$, as inferred from CMB observations, the chemical potential imprinted at the epoch of energy injection or loss can be
approximated analytically~
\cite{Hu:1993gc,Hu:1992dc,footnote1} 
as
\begin{equation}
 \mu=0.182 \left[\frac{t_{\rm cool}}{5.28 \times 10^6~{\rm s}}\right]^{1/2}
 \left[\frac{\Delta \rho_\gamma/n_\gamma}{6.75 \times 10^{-7}~{\rm GeV}}\right]\frac{\exp[{-(\tau_{\rm dC}/t_{\rm cool})^{5/4}}]}{0.293}~~,
 \label{distortion2}
\end{equation}
with
\begin{equation}
 \tau_{\rm dC}=6.22
\times10^6~{\rm s} \nonumber\\
\hspace{-3pt}
\times
\left[\frac{T_0}{2.73~{\rm K}}\right]^{
\hspace{-2pt}
-12/5}
\hspace{-3pt}
 \left[\frac{\Omega_b h^2}{0.0226}\right]^{
\hspace{-2pt}
4/5}
\hspace{-3pt}
 \left[\frac{1-Y_p/2}{0.877}\right]^{
\hspace{-2pt}
4/5}~~,
 \label{distortion3}
\end{equation}
where 
$t_{\rm cool}$ is the time of the photon cooling,
$\Delta \rho_\gamma$ is the difference between photon energy densities before and after the cooling,
$n_\gamma$ is the photon number density at the photon cooling epoch, $T_0$ is the present CMB temperature,
and
$Y_p$ is the primordial He abundance.

When energy transfers from photons to another particle, here assumed
to be a spinless boson, up to a factor of 1/3 of the initial photon
energy can be exchanged. This factor is derived assuming
a complete thermalization of photons and bosons with statistical degrees
of freedom of two and one, respectively. The cosmic time at which
photon cooling occurs should be much earlier than $\tau_{\rm dC}$.
Otherwise the amount of energy transfer should be diminished in order
to avoid an unrealistically large relic chemical potential. However,
in the original axion BEC model by Erken {\it et al}.~\cite{2012PhRvD..85f3520E}, the photon cooling occurs after the
temperature of the universe decreases to $\lesssim$ 500 eV.
If the cooling of photons through a gravitational interaction of axions
is effective enough, as supposed in Ref. \cite{2012PhRvD..85f3520E}
(case B), then the formation of an axion BEC followed by thermalization with
baryons and photons occurs after the cosmic temperature decreases to
$T_{\rm BEC} \sim 500~{\rm eV} X_a [f_a/(10^{12}~{\rm GeV})]^{1/2}$.
In this expression $X_a \lesssim 10$ is a factor related to the axion number
density at a critical time $t_1$ corresponding to the inverse mass of the axion, 
$t_1=m(t_1)^{-1}$, and $f_a$ is the axion decay constant.
The $f_a$ value is constrained to be $f_a \lesssim 10^{12}$GeV to avoid an axion energy density that is too large
\cite{1983PhLB..120..127P,1983PhLB..120..133A,1983PhLB..120..137D}.
The photon cooling if any, therefore, must occur at a temperature lower
than 1 keV.
This situation, however, is completely excluded by the very small
upper limit to $\mu$ consistent with the CMB black body
spectrum~\cite{2012PhRvD..85f3520E} unless both $X_a$
and $f_a$ are at their maximum allowed values. In Eq. (\ref{distortion2}),
numerical values correspond to the case that an energy transfer from
the CBR occurs at $T=500$ eV by as much as 1/3 of the total CBR
energy.

For late energy transfer at $z<10^5$, Compton scattering produces
little effect and cannot reestablish a Bose-Einstein spectrum. The
distorted spectrum can then be described by the Compton parameter $y$. There
is a relation between $y$ and the amount of the energy transfer, $\Delta \rho_\gamma/\rho_{\rm CBR}=4y$, where $\Delta \rho_\gamma$ and $\rho_{\rm CBR}$ are the
total transferred energy density and the CBR energy density, respectively. If 1/3 of the initial photon energy is transferred to another particle during the cooling, 
then the ratio of the energy injected into or extracted from the CBR energy
per comoving volume is
\begin{eqnarray}
 \frac{\Delta \rho_\gamma}{\rho_{\rm CBR}}=1/2.
 \label{distortion4}
\end{eqnarray}
So, the Compton $y$ value becomes $y=1/8$.

The CMB spectrum has been well measured. The limit on the chemical potential has been deduced to be $|\mu|<9\times 10^{-5}$ from an analysis of data from the Far-InfraRed Absolute Spectrophotometer on board the COsmic Background Explorer \cite{Fixsen:1996nj}. The Compton $y$-parameter is, on the other hand, constrained to be $|y|<1\times 10^{-4}$ based upon an updated constraint from the second generation of the Absolute Radiometer for Cosmology, Astrophysics, and Diffuse Emission \cite{footnote2}
utilizing a better fitting procedure~\cite{2011ApJ...734....6S}. The baryon density parameter is determined from the analysis of the Wilkinson Microwave Anisotropy Probe (WMAP) \cite{Spergel:2003cb,Spergel:2006hy,2011ApJS..192...16L,Hinshaw:2012aka}
to be $\Omega_b h^2\sim 0.02264$ with $h\sim 0.700$~\cite{Hinshaw:2012aka} for the $\Lambda$CDM model (WMAP only) from the WMAP 9yr data. Since a large amount of energy transfer is excluded by the $\mu$ and $y$ limits, energy transfer is allowed only when the transfer epoch occurs sufficiently early, i.e., $t_{\rm cool}<7.04 \times 10^5$ s.
\section{\label{sec:results}Results and Discussions}
\begin{figure}
\includegraphics[width=1.0\textwidth]{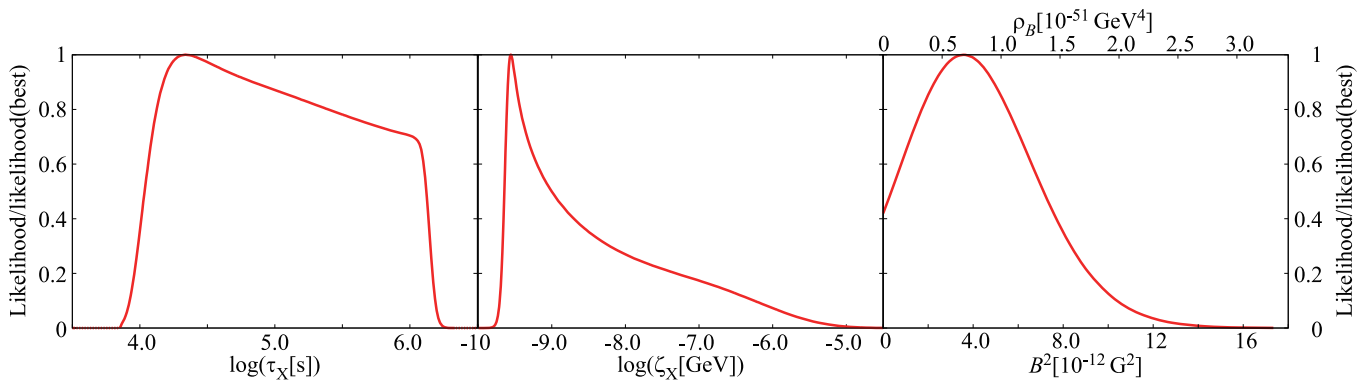}
\caption{\label{fig1}
Likelihood functions for $\tau_\mathrm{X}$ (left panel), $\zeta_\mathrm{X}$ (middle panel), and $B^2 \propto \rho_B$ (right panel) normalized to their maximum likelihood values.
} 
\end{figure}

The combined constraints on the $X$ particle and the PMF parameters in the new hybrid model introduced here were deduced in a maximum 
likelihood analysis. This leads to the limits on $\tau_\mathrm{X}$, $\zeta_\mathrm{X}$, and $B^2\propto \rho_B$ listed in Table \ref{table_cons}.
 Here $B$ is the comoving and scale-invariant strength of the PMF \cite{2012PhRvD..86l3006Y,2014PhRvD..89j3528Y}.
Figure \ref{fig1} also shows the probability distributions of $\tau_\mathrm{X}$, $\zeta_\mathrm{X}$, and $B^2\propto \rho_B$.
The red curves in this figure show the probability distributions. 
The obtained constraints on $\zeta_\mathrm{X}$ and $\tau_\mathrm{X}$ are
\begin{eqnarray}
4.06 < \log_{10}(\tau_X\mathrm{[sec]}) < 6.10~(2\sigma,~\mathrm{95\% ~C.L.})~~, \label{tau}\\
-9.70 < \log_{10}(\zeta_X\mathrm{[GeV]}) < -6.23~(2\sigma,~\mathrm{95\% ~C.L.})~~.\label{zeta}
\end{eqnarray}

We also find that the hybrid model with a PMF gives a better likelihood than without a PMF, and the best fit of the PMF energy density is 
\begin{eqnarray}
\rho_\mathrm{PMF}(a) = 
\frac{\rho_\mathrm{PMF} (a_0)}{a^4} = 6.82 \times 10^{-52} a^{-4}~ \mathrm{GeV}^4~(\mathrm{the~best~fit}),
\label{best_rho_PMF}
\end{eqnarray}
where $a$ is the scale factor.
This best fit value corresponds to 
\begin{eqnarray}
B(a) = 1.89 ~ a^{-2}~\mu\mathrm{G}~ (\mathrm{the~best~fit}).
\end{eqnarray}
Here, we use $\rho_B = B^2/8\pi = 1.9084\times 10^{-40}\times (B/\mathrm{G})^2 \mathrm{GeV}^4$.
However, we obtain only an upper bound on the PMF energy density at 2 $\sigma$ confidence level,
\begin{eqnarray}
\rho_\mathrm{PMF}(a) = 
\frac{\rho_\mathrm{PMF} (a_0)}{a^4} < 1.45 \times 10^{-51} a^{-4}~ \mathrm{GeV}^4~(2\sigma,~\mathrm{95\% ~C.L.}).
\label{rho_PMF}
\end{eqnarray}
This limit corresponds to 
\begin{eqnarray}
B(a) < 3.05 ~ a^{-2}~\mu\mathrm{G}~ (2\sigma,~\mathrm{95\% ~C.L.}).\label{amp_PMF}
\end{eqnarray}

The constrained magnetic energy density in this paper is higher than that deduced in previous studies \cite{2012arXiv1204.6164K,2012PhRvD..86l3006Y}. The reason is as follows.
For simplicity, we assume that the photon-cooling epoch after BBN is followed by a complete thermalization of the cosmic background radiation.
After cooling, the baryon-to-photon ratio is then higher than in the case without cooling.
Also, the ratio of energy densities of the magnetic field to the photons is higher by a factor of $(\eta/\eta_{\rm BBN})^{4/3}=3/2$. Consequently, the comoving PMF amplitude $B$ after the epoch of photon cooling is a factor of $(3/2)^{1/2}$ larger than in the case without cooling.

\begin{figure}
\includegraphics[width=1.0\textwidth]{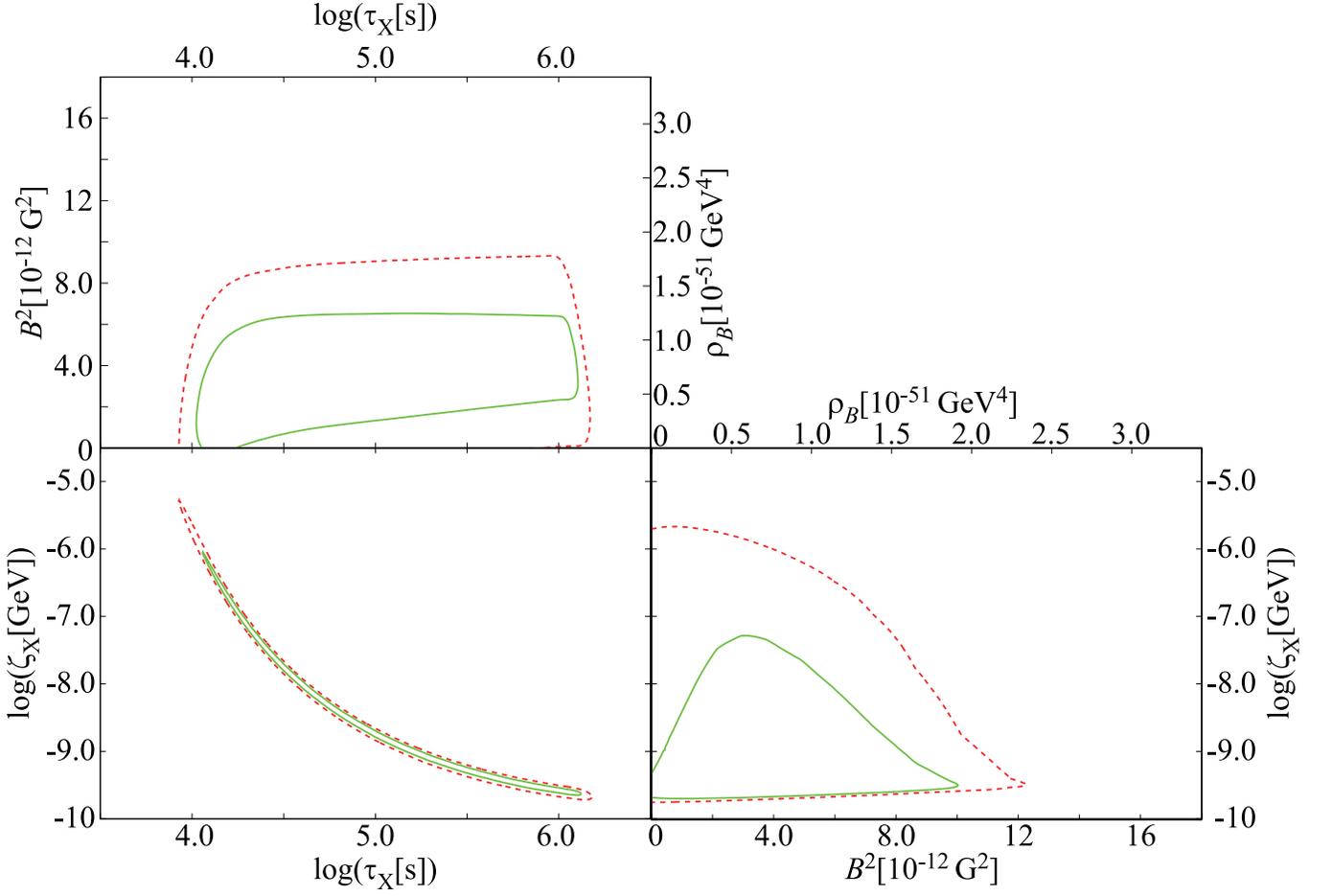}
\caption{\label{fig2}
Contours of the 1$\sigma$ (68\%; solid curves) and 2$\sigma$ (95\%; dotted curves) confidence limits on various planes for parameters characterizing the $X$ particles and the PMF.
} 
\end{figure}

\begin{figure}
\includegraphics[width=1.0\textwidth]{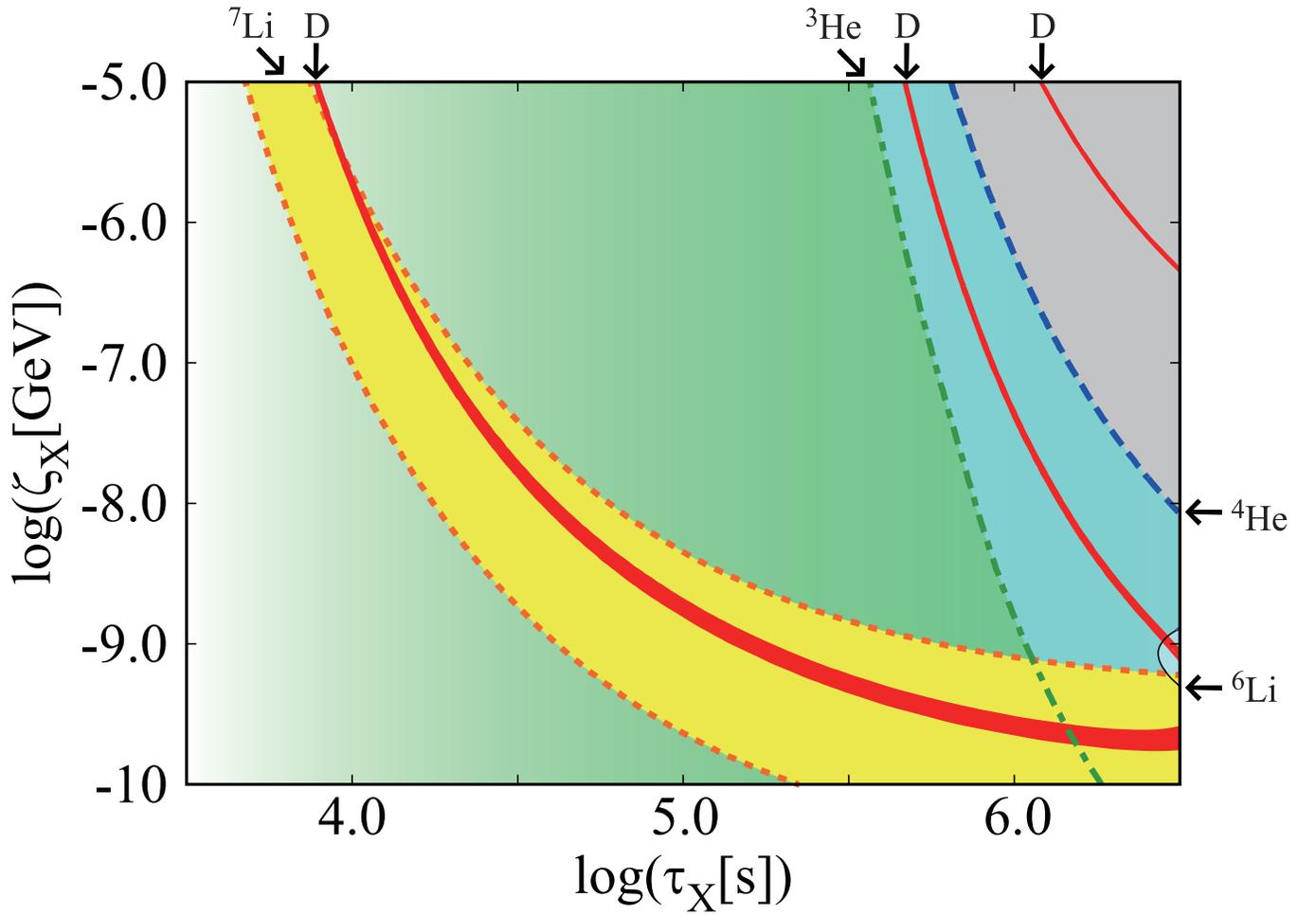}
\caption{\label{fig3}
Allowed regions in the ($\tau_X$, $\zeta_X$) plane for the hybrid model with a PMF for $\eta = 4.57 \times 10^{-10} $ and $B=1.89~\mu$G. The left sides of these curves denote the allowed regions derived from observational limits on the primordial elemental abundances. The narrow dark band and the region bounded by the dotted curves (color online: red and yellow regions) show the 2$\sigma$ (95\%) confidence limits determined from the observed abundances of D and $^7$Li, respectively. Dot-dashed, dashed and thin solid curves (color online: green, blue and black curves) are the 2$\sigma$ (95\%) confidence limits determined from the upper limits on the $^3$He, $^4$He and $^6$Li abundances.
} 
\end{figure}

Figure \ref{fig2} shows 
contours of 1 and 2 $\sigma$ confidence limits
on various planes of the $X$ particle and PMF parameters (i.e.,~$\tau_\mathrm{X}$ vs. $\zeta_\mathrm{X}$, $\rho_B$ vs. $\zeta_\mathrm{X}$ and $\rho_B$ vs $\tau_\mathrm{X}$).
he solid and dotted (color online: green and red) contours show the 1 $\sigma$(68\%) and 2 $\sigma$(95\%) confidence limits. 
In the BBN model including the decaying $X$ particle without a PMF, parameters of the $X$ particle ($\tau_X$, $\zeta_X$) are constrained most strongly from the observational limits on the abundances of D and $^7$Li, and parameters in the allowed region are strongly degenerate \cite{2013PhLB..718..704K}. Our constraints on these parameters are shown in Fig. \ref{fig3}. These are consistent with those of previous work \cite{2013PhLB..718..704K}.

The $X$ particle parameters $(\tau_\mathrm{X}, \zeta_\mathrm{X})$ are not strongly constrained by the $^4$He abundance.
On the other hand, the PMF energy density is mainly constrained by the $^4$He abundance. 
No degeneracy is then found between the parameters of the PMF and the $X$ particle in the allowed region of this hybrid BBN model with a PMF as shown in Fig.\ref{fig2}.

Finally, it is worthwhile to mention that one should compute both BBN and the CMB and compare these theoretical predictions with observations of the light element abundances and the CMB spectra simultaneously. This is because constraints on $N_\mathrm{eff}$ from the observed CMB data sets alone generally do not take consistency with the light element abundances into account except for $^4$He. As a result, the $N_\mathrm{eff}$ value which is consistent with observational light element abundance constraints is different from those constrained from the observed CMB data (Table \ref{tabel2}).

In summary, we have calculated BBN taking into account three possible cosmological extensions of the standard BBN.  These include photon cooling, the radiative decay of $X$ particles, and the possible existence of a PMF. In particular, we consider the possible combination of all three paradigms simultaneously in a new hybrid model. We then utilized a maximum likelihood analysis to deduce constraints on the parameters characterizing the $X$ particles ($\tau_\mathrm{X}$, $\zeta_\mathrm{X}$) and the energy density of the PMF ($\rho_B = B^2/8\pi$) from the observed abundances of light elements up to Li.

As a result, we obtained ranges for the $X$-particle parameters given by Eqs. (\ref{tau}) and (\ref{zeta}).
We also find that the hybrid model with a PMF gives the better likelihood than that without a PMF. The best fit and 2 $\sigma$ upper bound on the energy density of the PMF are given by Eqs. (\ref{best_rho_PMF}) and (\ref{rho_PMF}).

Since the parameters of the $X$ particle are mainly constrained by the D and $^7$Li abundances, while the energy density of the PMF is constrained by the $^4$He abundance, we found that there are no significant degeneracies between parameters of the PMF and the $X$ particle as shown in Fig.\ref{fig2}.
The effective number of neutrino species $N_\mathrm{eff}$ within our new hybrid model is based only upon consistency with the observed light elements.
Hence, our constraint differs from that based upon the CMB.
In a subsequent work we will report on a computation of the constraints from
both BBN and the CMB simultaneously to deduce a new limit on $N_\mathrm{eff}$.

\begin{table}[!t]
\caption{Agreement with observed light element abundances for the five
models considered here.}
\begin{minipage}{\textwidth}
\begin{tabular}{c|c|c|c|c|c|c}
\hline
\multicolumn{2}{c|}{Model}                  & SBBN
  & $\gamma$-cooling & $X$ particle & $B$ field   &
\shortstack{$\gamma$-cooling+$X$ particle \\ + $B$ field} \\
\hline											
\multicolumn{2}{c|}{$\eta$ ($\times 10^{10}$)}        & $6.19
\pm 0.14$ & $4.57 \pm 0.10$ \footnotemark[1] &
\multicolumn{2}{c|}{$6.19 \pm 0.14$} & $4.57 \pm 0.10$
\footnotemark[1] \\
\hline											
Nuclide    & Observation & \multicolumn{5}{c}{} \\						
\hline											
$Y_{\rm p}$ ($^4$He)    & 0.2345--0.2777\footnotemark[2] &
\checkmark & \checkmark    & \checkmark  & \checkmark  &
\checkmark \\
D/H ($\times 10^5$)     & 2.37--2.85           &
\checkmark & -        & \checkmark  & \checkmark  &
\checkmark \\
$^3$He/H ($\times 10^5$)  & 0--3.1             &
\checkmark & \checkmark    & \checkmark  & \checkmark  &
\checkmark \\
$^7$Li/H ($\times 10^{10}$) & 1.06--2.35 \footnotemark[3]  & -
  & ?        & -      & -       & \checkmark \\
\hline									 		  		
$^6$Li/H ($\times 10^{12}$) & 0--9.5 \footnotemark[4]    &
\checkmark & \checkmark    & \checkmark (high)     & \checkmark  & \checkmark (high)
  \\
\end{tabular}
\footnotetext[1]{$\eta_{\rm BBN}=(2/3)^{3/4}\eta_{\rm
WMAP}$~\cite{2012PhRvD..85f3520E,Erken:2011vv}.}
\footnotetext[2]{Conservative limit \cite{Aver:2010wq}.}
\footnotetext[3]{Spite plateau value~\cite{Sbordone2010} in metal-poor
halo stars.}
\footnotetext[4]{2$\sigma$ upper limit from a spectral analysis of
metal-poor halo stars \cite{2013AA...554A..96L}.}
\end{minipage}
\label{tab:models}
\end{table}

\clearpage

\begin{table}
\begin{center}
\caption{Constraints on $N_\mathrm{eff}$ and $\eta$ from the CMB. \label{tabel2}}
\begin{tabular}{ccccc}
\tableline\tableline
\multicolumn{5}{c}{Fixed $N_\mathrm{eff}=3.046$ with WMAP only}\\
&
WMAP&
W+h0&
W+bao&
W+h0+bao\\
\tableline
$\eta$  &
$6.19~\pm~0.14$   &
$6.26~\pm~0.13$   &
$6.15~\pm~0.12$   &
$6.20~\pm~0.12$   \\
\tableline\tableline
\multicolumn{5}{c}{Free $N_\mathrm{eff}$ with WMAP only}\\
\tableline
$N_\mathrm{eff}$  &
$> 1.7 (2\sigma)$   &
$3.96^{+0.75}_{-0.74}$   &
$4.9^{+2.4}_{-2.2}$   &
$4.23~\pm~0.59$   \\
$\eta$  &
$6.20~\pm~0.14$   &
$6.20~\pm~0.13$   &
$6.16~\pm~0.12$   &
$6.16~\pm~0.12$   \\
\tableline
\tableline\tableline
\multicolumn{5}{c}{Fixed $N_\mathrm{eff}=3.046$ with WMAP + ACT + SPT + SNLS3}\\
&
WMAP&
W+h0&
W+bao&
W+h0+bao\\
\tableline
$\eta$  &
$6.12~\pm~0.10$   &
$6.146^{+0.097}_{-0.096}$   &
$6.055^{+0.091}_{-0.090}$   &
$6.085~\pm~0.09$   \\
\tableline\tableline
\multicolumn{5}{c}{Free $N_\mathrm{eff}$ with WMAP + ACT + SPT + SNLS3}\\
\tableline
$N_\mathrm{eff}$  &
$3.97~\pm~0.66$   &
$3.66^{+0.40}_{-0.39}$   &
$3.61~\pm~0.60$   &
$3.83~\pm~0.40$   \\
$\eta$  &
$6.23~\pm~0.13$   &
$6.186~\pm~0.099$   &
$6.12~\pm~0.11$   &
$6.148^{+0.097}_{-0.093}$   \\
\tableline
\end{tabular}
\end{center}
\end{table}

\begin{table}
\caption{Constraints on the parameters of $X$ particle and the PMF amplitude.
\label{table_cons}}
\begin{tabular}{cccccc} 
\hline
\multicolumn{1}{c}{constrained values (95\% C.L.)}&
\multicolumn{1}{c}{best} \\
\hline
$ 4.06 < \log_{10}(\tau_X\mathrm{[s]}) < 6.10$ &
4.34 \\
$ -9.70 < \log_{10}(\zeta_X\mathrm{[GeV]}) < -6.23$&
-9.55 \\
 $B < 3.05 ~\mu\mathrm{G}$ &
$1.89~\mu\mathrm{G}$ \\
\hline
\multicolumn{2}{l}{Confidence intervals (2$\sigma$,95\% C.L.) and upper bounds ($2\sigma$,95\% C.L.) on strengths of PMFs.} \\
\hline
\end{tabular}
\end{table}
\begin{acknowledgments}
This work was supported in part by Grants-in-Aid for Scientific Research of JSPS (D.G.Y. by Grant No. 25871055, T. K. by Grants No. 20105004, No. 24340060). 
 Work at the University of Notre Dame (G.J.M.) is supported by  
the U.S. Department of Energy under Nuclear Theory Grant No. DE-FG02-95-ER40934.
Work (M.K.C.) is supported by the National Research Foundation of Korea (2012M7A1A2055605,2011-0015467).
\end{acknowledgments}
\bibliographystyle{apsrev}

\end{document}